\documentclass[12pt]{article}

\usepackage{graphicx}
\usepackage{color} 
\usepackage{hyperref}
\usepackage{amsmath,amssymb,mathrsfs}
\usepackage{psfrag}
\usepackage{graphicx}
\usepackage{graphics}
\usepackage{epsfig}
\usepackage{bm}
\usepackage{verbatim,color,ulem}
\usepackage{braket}
\usepackage{verbatim}
\usepackage{placeins}

\usepackage{titlesec}
\titleformat{\subsection}[runin]{\bfseries}{}{0.0em}{}

\newcommand{\beq}{\begin{equation}}
\newcommand{\eeq}{\end{equation}}

\def\ket#1{|#1\rangle}

\newcommand{\be}{\begin{equation}}
\newcommand{\ee}{\end{equation}}

\def\correspondingauthor{\footnote{Correspondence to \texttt{macri@fisica.ufrn.br}.}}

\title{Equation of state and self-bound droplet in Rabi-coupled Bose mixtures}

\author{A. Cappellaro$^{1}$, T. Macr\`\i$^{2}\correspondingauthor{}$, G. F. Bertacco$^1$ and L. Salasnich$^{1,3}$}
\date{\today}
\begin{document}

\maketitle
\begin{center}
$^1$ Dipartimento di Fisica e Astronomia "Galileo Galilei'', 
Universit\`a di Padova, via Marzolo 8, 35131 Padova, Italy
\\
$^2$ Departamento de F\'isica Teorica e Experimental, Universidade 
Federal do Rio Grande do Norte, 
and International Institute of Physics, 59070-405, Natal-RN, Brazil
\\
$^3$ Istituto Nazionale di Ottica (INO) del Consiglio Nazionale 
delle Ricerche (CNR), via Nello Carrara 1, 50019 Sesto Fiorentino, Italy
\end{center}
\begin{abstract}
Laser induced transitions between internal states of atoms have been playing a fundamental
role to manipulate atomic clouds for many decades. 
In absence of interactions each atom behaves independently and their coherent 
quantum dynamics is described by the Rabi model. 
Since the experimental observation of Bose condensation in dilute gases,
static and dynamical properties of multicomponent quantum gases have been
extensively investigated.
Moreover, at very low temperatures quantum fluctuations crucially affect the equation of state
of many-body systems. Here we study the effects of quantum fluctuations on a 
Rabi-coupled two-component Bose gas of interacting alkali atoms. 
The divergent zero-point energy of gapless and 
gapped elementary excitations of the uniform system is properly regularized 
obtaining a meaningful analytical expression for the 
beyond-mean-field equation of state. 
In the case of attractive inter-particle interaction 
we show that the quantum pressure arising from Gaussian fluctuations 
can prevent the collapse of the mixture with the creation of a 
self-bound droplet. 
We characterize the droplet phase and 
discover an energetic instability above a critical Rabi frequency provoking
the evaporation of the droplet. 
Finally, we suggest an experiment  to observe 
such quantum droplets using Rabi-coupled internal states of $^{39}$K atoms.
\end{abstract}

\flushbottom

\thispagestyle{empty}

\section*{Introduction}
In atomic physics, laser beams can stimulate 
transitions among different hyperfine states. 
For bosonic atoms at temperatures below the transition to the superfluid phase, 
coupling of hyperfine states offers the possibility 
to address fascinating phenomena
such as the internal Josephson effect \cite{Leggett2001,Hall1998,Hall1998_2} emulating a space dependent
double well potential, analogues of the Hawking 
radiation \cite{Carusotto2017,Larre2013}, non-abelian gauge 
potentials \cite{Dalibard2011} like magnetic monopoles 
\cite{Mottonen2009, Mottonen2014}, Rashba spin-orbit 
coupling \cite{Merkl2008, Song2009,Li2012,Martone2012} 
or they can be used for applications to quantum metrology \cite{Gross11,Lucke11,Macri16}
and for  the quantum simulation of spin models with short or long-range interactions 
\cite{Fukuhara11,Schauss14,Zeiher16,Labuhn16}. 

In this article we study the effects of a Rabi coupling on a two-component 
Bose mixture deriving the corresponding beyond-mean-field equation of state. 
To achieve this result we perform a non-trivial regularization of Gaussian 
fluctuations, which have a divergent zero-point energy 
due to both gapless and gapped elementary excitations. 
In particular, we obtain a meaningful analytical formula for the ground-state 
energy of the Bose mixture as a function of Rabi coupling 
and scattering lengths. Setting the Rabi frequency to zero in our formula 
one recovers Larsen's equation of state \cite{Larsen1963}. 
In the case of attractive inter-particle 
interaction we investigate the conditions for the formation of 
a self-bound droplet finding that its density profile and collective 
oscillations crucially depend on the interplay between Rabi 
coupling and interaction strengths.   
A similar equation of state, albeit in absence of internal coupling,
 has been recently used by Petrov 
\cite{Petrov2015,Petrov2016}. He shows that, in the case of negative 
inter-component scattering length, quantum fluctuations can arrest the collapse 
of the mixture inducing the formation of a stable self-bound droplet. 
In a different context, the stabilization induced by quantum 
fluctuations has been found also in dipolar Bose-Einstein condensate, 
both in trapped configuration \cite{Kadau16,Wachtler2016_fil} 
and in free space \cite{Schmitt16,Blakie2016_R, Blakie2016,Wachtler2016}. 

Remarkably, we find that above a critical Rabi 
frequency the self-bound droplet evaporates into a uniform configuration 
of zero density. Finally, we analyze the most favorable conditions to obtain 
a stable self-bound droplet made of $^{39}$K atoms in two Rabi-coupled 
hyperfine states.

\section*{Results}

\subsection*{Microscopic theory for Rabi-coupled mixtures.}
We consider a Bose gas with two relevant hyperfine states in a volume $L^3$, 
at temperature $T$ and with chemical potential $\mu$. In addition to the usual intra- 
and inter-state contact interactions, transitions between the two states are induced 
by an an external coherent Rabi coupling of frequency $\omega_R$. 
We adopt the path integral formalism, where each component is described 
by a complex bosonic field $\psi_i$ ($i = 1,2$). Given the spinor 
$\Psi = (\psi_1,\psi_2)^T$
\cite{armaitis2015,schakel2008,Stoof2009Ultracold}, 
the partition function of the system reads: 
\begin{equation}
\mathcal{Z} = \int \mathcal{D}[\Psi,\bar{\Psi}] 
\exp\bigg( - \frac{1}{\hbar} S[\Psi,\bar{\Psi}]\bigg) \;,
\label{def:Z}
\end{equation}
where the Euclidean action $S[\Psi,\bar{\Psi}]$ is given by
\begin{equation}
\begin{aligned}
S\big[\Psi,\bar{\Psi}\big] & =  
\int_0^{\beta\hbar}d\tau\int_{L^3} d^3\mathbf{r} \bigg\lbrace \sum_{i=1,2}
\bigg[ \psi_i^* \left(\hbar\partial_{\tau} -\frac{\hbar^2}{2m}\nabla^2 - \mu\right)
\psi_i + \frac{1}{2}g_{ii} 
|\psi_i|^4 \bigg] \\
& \qquad \qquad \qquad \qquad + g_{12}|\psi_1|^2 |\psi_2|^2 - \hbar\omega_R(\psi^*_1\psi_2 + 
\psi_2^* \psi_1)\bigg\rbrace ,\\
\end{aligned}
\label{euclidean action}
\end{equation}
with $\beta \equiv 1/(k_BT)$
and 
$g_{ij} =  4\pi\hbar^2 a_{ij}/m$ being $a_{ij}$ the 
scattering length for collisions between component $i$ and 
component $j$ (specifically $a_{11}$, $a_{22}$, and $a_{12}$). 
All relevant thermodynamical quantities can be derived from
 the grand potential $\Omega = -\frac{1}{\beta}\text{ln}\big(\mathcal{Z}\big)\;$.
We work in the superfluid phase, where a $\text{U}(1)$
gauge symmetry of each bosonic component is spontaneously broken. 
The presence of the Rabi coupling in the Euclidean action in equation
\eqref{euclidean action}
implies that only the total number of atoms is conserved.
We can then set $\psi_i(\mathbf{r},\tau) = v_i + \eta_i(\mathbf{r},\tau)\;$,
where $v_i$ are the uniform order parameters of the two-component 
Bose-Einstein condensate, and $\eta_i({\bf r},\tau)$ are the fluctuation 
fields above the condensate. The mean-field plus gaussian approximation
is obtained by expanding equation \eqref{euclidean action} up to the 
second order in $\eta_i({\bf r},\tau)$ and $\eta_i^*({\bf r},\tau)$. 
The corresponding beyond-mean-field grand potential is 
then given by \cite{Stoof2009Ultracold,andersen2004}
\begin{equation}
\Omega(\mu, v_1,v_2) = \Omega_0(\mu,v_1,v_2) + \Omega_g(\mu,v_1,v_2) \; ,
\label{luca's recipe}
\end{equation}
where 
\begin{equation}
\begin{aligned}
\Omega_0 (\mu,v_1,v_2) = L^3\bigg[\sum_{i=1,2} \bigg(-\mu v_i^2 + 
\frac{1}{2}g_{ii} v_i^4\bigg) + g_{12}v_1^2v_2^2 -2\hbar\omega_R v_1v_2\bigg] \;
\end{aligned}
\label{omega0_gen}
\end{equation}
is the mean-field grand potential, while $\Omega_g(\mu,v_1,v_2)$
is the grand potential of Gaussian quantum and thermal fluctuations.

\begin{figure}
\centering
\includegraphics[width=\textwidth,clip=]{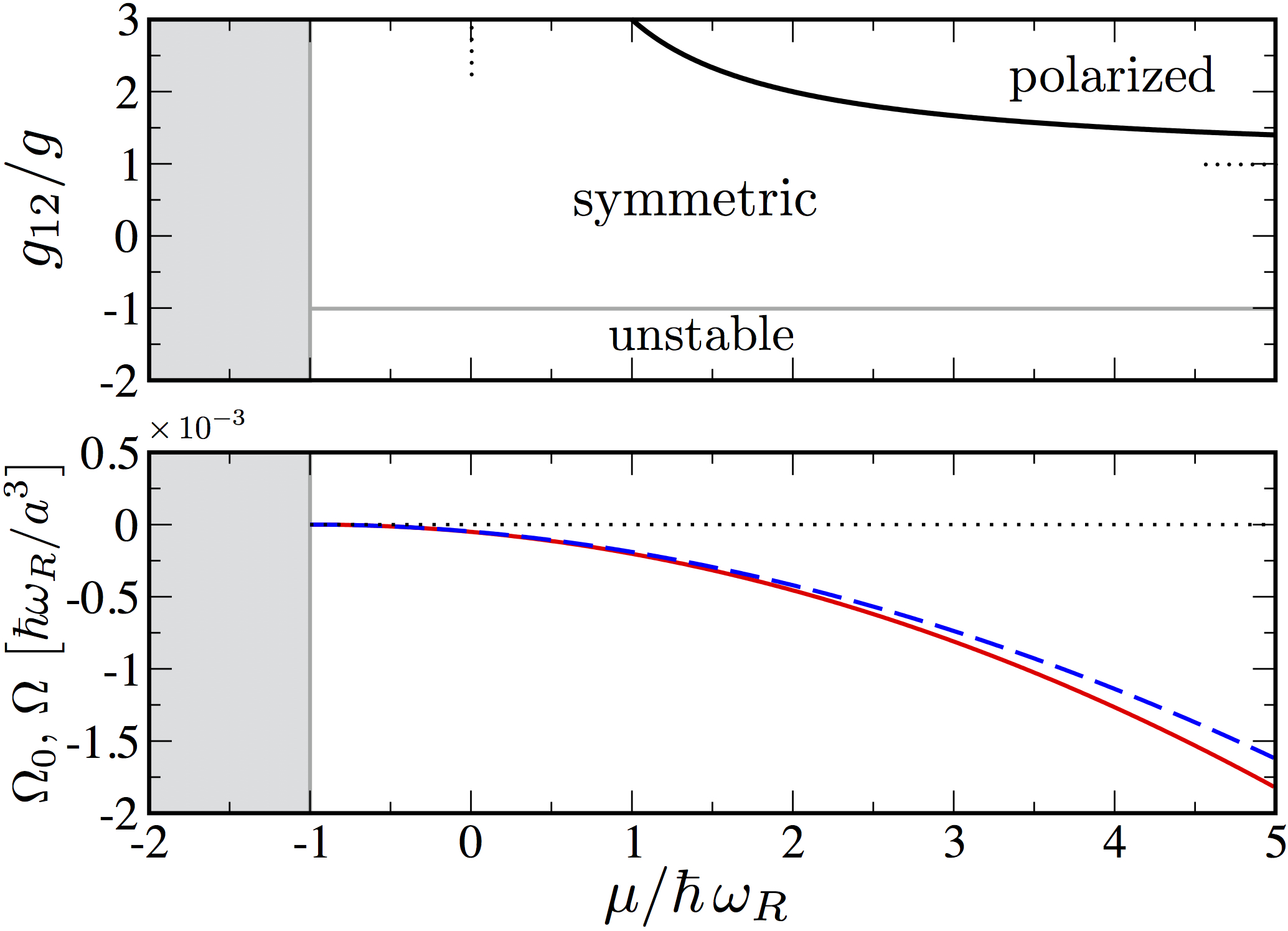}
\caption{{\bf Grand canonical phase diagram and grand potential.}
(Top) Mean field phase diagram based on the grand potential $\Omega_0(\mu,\nu_1,\nu_2)$ 
of equation (\ref{omega0_gen}).
In the symmetric ground state the two components appear with the same particle density $|\nu_1|^2=|\nu_2|^2$,
whereas in the polarized phase densities are unequal. 
Dotted lines represent the asymptotic phase boundaries of the polarized region for large $g_{12}/g$ and 
$\mu/\hbar\, \omega_R$ ratios respectively.
For $g_{12}/g<-1$ the symmetric solution is unstable in the 
\textit{thermodynamic limit}. The grey region for $\mu<\hbar\, \omega_R$ corresponds to the trivial solution
$|\nu_1|^2=|\nu_2|^2=0$.
(Bottom) Grand potential $\Omega(\mu)$ (dashed blue line) with the inclusion of gaussian fluctuations of
equation (\ref{oneloop_F}) and its mean field approximation $\Omega_0(\mu)$ (red solid line) of 
equation (\ref{omega0 mf}) as a function of the chemical
potential $\mu$ for $g_{12}=0.9\, g$ within the symmetric phase. 
}
\label{fig1}
\end{figure}

In our scheme, the Bose-Einstein order parameters $v_i$ satisfy the saddle-point 
equations $\partial \Omega_0\big(\mu, v_1,v_2 \big)/\partial v_i = 0\;$, leading to
coupled equations for the uniform and 
constant fields $v_1$ and $v_2$:
\begin{equation}
(g_{ii} v_i^2 + g_{ij} v_j^2)v_i - \hbar\omega_R v_{j} = \mu v_i
\label{saddle-point}
\end{equation}
with $i=1,2$ and $j\neq i$. 
The analysis of the minima of $\Omega_0\big(\mu, v_1,v_2 \big)$ at the solution of equations (\ref{saddle-point})
leads to the mean field phase diagram of Fig. \ref{fig1}\, a which is obtained for the case of equal
intra-component {\it repulsive} interaction strength  $g_{11} = g_{22} \equiv g$.  
%%%%%%%%%%%%%%%%%%%%%%%%%%%%%%
%%%%%%%%%%%%%%%%%%%%%%%%%%%%%%
One finds a symmetric configuration where the two internal states are equally populated,
a polarized phase with non-zero population imbalance, and a unstable phase 
when the attractive inter-state interaction overcomes the intra-state repulsion $g_{12}<-g$.
\cite{abad2013,lellouch2013,search2001,Tommasini2003}.

In the rest of this article we focus on the symmetric ground 
state, where $v_1 = v_2 \equiv v/\sqrt{2}$ and equal intra-component interaction. The corresponding 
mean-field grand potential $\Omega_0(\mu, v)$ 
is then given by 
\begin{equation}
\frac{\Omega_0(\mu,v)}{L^3} = -\mu v^2 + \frac{1}{4}(g + g_{12}) v^4 
-\hbar \omega_R v^2 \;.
\label{omega0_v}
\end{equation}
By solving equation \eqref{saddle-point} in the case of symmetric ground-state, 
we get the crucial relation between the order parameter and the chemical potential:
$v^2 = 2(\mu + \hbar\omega_R)/(g + g_{12})$.
In this case, equation \eqref{omega0_v} reduces to 
\begin{equation}
\frac{\Omega_0(\mu)}{L^3} = - \frac{(\mu+\hbar\omega_R)^2}{g+g_{12}}.
\label{omega0 mf}
\end{equation}

\subsubsection*{Gaussian Fluctuations.} To compute $\Omega_g(\mu,v)$ for the symmetric ground state and
for equal interaction strengths, we consider 
the quadratic terms in $\eta_i$ and $\eta_i^*$ of equation \eqref{euclidean action}. 
In reciprocal Fourier space one finds 
\begin{equation}
S_2\big[\mathbf{\eta}(\mathbf{q},\omega_n), \bar{\mathbf{\eta}}
(\mathbf{q},\omega_n) \big] 
= -\frac{\hbar}{2}\sum_{\mathbf{q},\omega_n} \bar{\mathbf{\eta}}
(\mathbf{q},\omega_n) \ 
\mathbb{M}(\mathbf{q},\omega_n) \ \mathbf{\eta}(\mathbf{q},\omega_n) \; .
\label{S2 fourier}
\end{equation}
Here $\lbrace \omega_n \rbrace_n$ are the bosonic Matsubara frequencies and $-\hbar \mathbb{M}(\mathbf{q},\omega_n)$
is the $4\times 4$ inverse of the fluctuations propagator, whose definition 
is reported in the Methods.
At zero temperature, the Gaussian grand potential corresponds to
the zero-point energy of bosonic excitations and it reads 
\cite{Stoof2009Ultracold, Salasnich2016}
\begin{equation} 
\Omega_g(\mu,v) = \frac{1}{2}\sum_{\mathbf{q}}\big[E_{\mathbf{q}}^{(+)}(\mu,v) 
+ E_{\mathbf{q}}^{(-)}(\mu,v)\big] \;,
\label{omega gaussian}
\end{equation} 
where $E^{(\pm)}_{\mathbf{q}}(\mu,v)$ is the spectrum 
of elementary excitations, which can be obtained by diagonalizing 
$-\hbar\big[\mathbb{I}\cdot \mathbb{M}(\mathbf{q},0)\big]$ 
\cite{armaitis2015, Stoof2009Ultracold,Fetter}. 
The diagonal blocks of $\mathbb{I}$ are two-by-two identity matrices $\mathbf{1}_2$, 
while the off-diagonal ones are the Pauli matrix $\sigma_z$. 
The eigenvalues are the two branches of the Bogoliubov spectrum:
\begin{equation}
E^{(+)}_{\mathbf{q}} = \sqrt{\frac{\hbar^2 q^2}{2m}\bigg[\frac{\hbar^2 q^2}{2m} 
+ 2\big(\mu +\hbar\omega_R\big)\bigg]}
\label{bogoliubov spectrum 1}
\end{equation}
and 
\begin{equation}
E^{(-)}_{\mathbf{q}} = \sqrt{\frac{\hbar^2 q^2}{2m}\bigg[\frac{\hbar^2 q^2}{2m}
+ 2\,A(\mu,\omega_R)\bigg] + B(\mu,\omega_R)}\;,
\label{bogoliubov spectrum 2}
\end{equation}
where we set $\epsilon=a_{12}/a$, with $a$ the intra-component scattering 
length and $a_{12}$ the inter-component scattering lengt,
$A  = (\mu+\hbar\omega_R)(1-\epsilon)/(1+\epsilon) + 2\hbar\omega_R$
and 
$B = 4\hbar\omega_R \left[(\mu+\hbar\omega_R)(1-\epsilon)/(1+\epsilon) 
+ \hbar\omega_R\right].$
In the continuum limit $\sum_\mathbf{q} \rightarrow L^3 
\int d^3\mathbf{q}/(2\pi)^3$, 
the zero-temperature Gaussian grand potential is ultraviolet divergent. 
We employ the convergence-factor regularization 
\cite{Stoof2009Ultracold,Diener2008,Salasnich2016} which generates proper 
counterterms in the zero-point energy completely removing the divergence. 
These counterterms can be determined by expanding 
the two branches of the Bogoliubov spectrum at high momenta.   
The zero-temperature beyond-mean-field grand potential is then 
given by equation \eqref{omega0 mf} plus the regularized zero-point energy, namely 
\begin{equation}
\begin{aligned}
\frac{\Omega(\mu)}{L^3} & = -\frac{(\mu + \hbar \omega_R)^2}{g+g_{12}} 
+ \frac{8}{15\pi^2}\bigg(\frac{m}{\hbar^2}\bigg)^{3/2}\big(\mu +
\hbar\omega_R\big)^{5/2} \\
& \qquad \qquad \qquad \qquad \qquad + \frac{A^{5/2}}{2\sqrt{2}\pi^2}\bigg(\frac{m}{\hbar^2}
\bigg)^{3/2} 
I(\mu,\omega_R, \epsilon) \;.
\end{aligned}
\label{oneloop_F}
\end{equation}
The function $I(\mu,\omega_R,\epsilon)$ is given by
\begin{equation}
\begin{aligned}
&I\big(\mu,\omega_R, \epsilon) = \int_0^{+\infty}dy\bigg[ \sqrt{y} 
\sqrt{y^2 +2y+\frac{B}{A^2}} -y^{3/2} - \sqrt{y} - 
\frac{\big(\frac{B}{A^2} -1\big)}{2\sqrt{y}} \bigg] \;.\\
\end{aligned}
\label{grancan_integralazzo}
\end{equation} 
In Fig. \ref{fig1}\,b we plot the grand potential $\Omega(\mu)$ of equation (\ref{grancan_integralazzo}),
including gaussian fluctuations, as a function of the chemical potential for $g_{12}=0.9\, g$. 
We compare it with the mean field approximation $\Omega_0(\mu)$ of equation (\ref{omega0 mf}).
The energy density of the system is $\mathcal{E} = E/L^3 = \Omega/L^3 + \mu n$ where
the number density $n$ is obtained via $n= {1\over L^3} {\partial \Omega \over \partial \mu} \; $.
In the limit of small Rabi-coupling, which is also the 
most relevant experimentally \cite{Carusotto2017} (see below), it is possible to get an analytical result for 
the energy density. By taking $E_B= \hbar^2/ma^2$ as energy unit 
(then $\hbar\omega_R= \bar{\omega}_R E_B$) and defining the {\it diluteness parameter}
$\bar{n}=na^3$, up to the linear term in $\bar{\omega}_R$, 
from equation (\ref{oneloop_F}) we obtain
the scaled energy density of the uniform bosonic mixture  
\begin{equation}
\begin{aligned}
\frac{\mathcal{E}}{E_B/a^3}& = \pi\big(1 + \epsilon\big)\bar{n}^2 
-\bar{\omega}_R\bar{n}+ \frac{8}{15\pi^2}\big[2\pi\bar{n}(1+\epsilon)\big]^{5/2}  \\
&\qquad \qquad +\frac{8}{15\pi^2}\big[2\pi \bar{n} (1-\epsilon)\big]^{5/2} 
+ \frac{14}{3\pi^2}\bar{\omega}_R \big[2\pi \bar{n}(1-\epsilon) 
\big]^{3/2} \,.
\end{aligned}
\label{canonical energy functional}
\end{equation} 
Notice that for $\bar{\omega}_R=0$ one recover the Larsen's 
zero-temperature equation of state \cite{Larsen1963}. 
From equation (\ref{canonical energy functional}) 
one finds that for $|\epsilon| > 1$ the uniform configuration 
is not stable. If $\epsilon > 1$, at the mean field level, one expects phase separation 
or population imbalance \cite{abad2013}. Instead, 
if $\epsilon < -1$ the term proportional 
to $[(1+\epsilon)\bar{n}]^{5/2}$ becomes imaginary and it gives rise 
to a dissipative dynamics. However, this dissipative 
term can be neglected if $\bar{n}$ is not too 
large (as well as other sources of losses like three-body recombination). 
The resulting {\it real} energy density 
displays a characteristic $\bar{n}^{5/2}$ dependence which 
competes with the negative mean-field contribution, opening the door 
to the possibility of observing a droplet phase for finite systems.
This stabilization mechanism based on quantum fluctuations has been proposed for the first time in 
two-component mixtures without Rabi coupling \cite{Petrov2015,Petrov2016} 
and recently applied to dipolar condensates \cite{Wachtler2016, Blakie2016,Blakie2016_R}.
For $\epsilon < -1$ the equilibrium density is obtained upon the 
minimization of the energy density in equation (\ref{canonical energy functional}) 
with respect to $\bar{n}$ neglecting the imaginary term: 
\begin{equation}
\bar{n}_{\pm} = \left( 
\frac{5\sqrt{\pi}\big|1+\epsilon \big|}{32\sqrt{2}(1+|\epsilon|)^{5/2}}
\bigg[1 \pm \sqrt{1 - \frac{1792\,\bar{\omega}_R}{15\pi^2}
\frac{(1+|\epsilon|)^4}{|1+\epsilon|^2}}\bigg] \right)^2 \; . 
\label{trova-n}
\end{equation}  
The solution $\bar{n}_{-}$ is a local maximum, 
while the equilibrium value is given by $\bar{n}_{+}$ 
which is a local minimum of the energy per particle. 
Moreover to obtain a real solution, Rabi frequency is limited by:  
$\bar{\omega}_R < \frac{15\pi^2}{1792}\frac{|1+\epsilon |^2}{(1+|\epsilon|)^4}$. 
For larger $\bar{\omega}_R$ there 
is only the absolute minimum with zero energy at $\bar{n}=0$. 

\subsection*{Droplet phase}
\begin{figure}
\centering
\includegraphics[width=\textwidth,clip=]{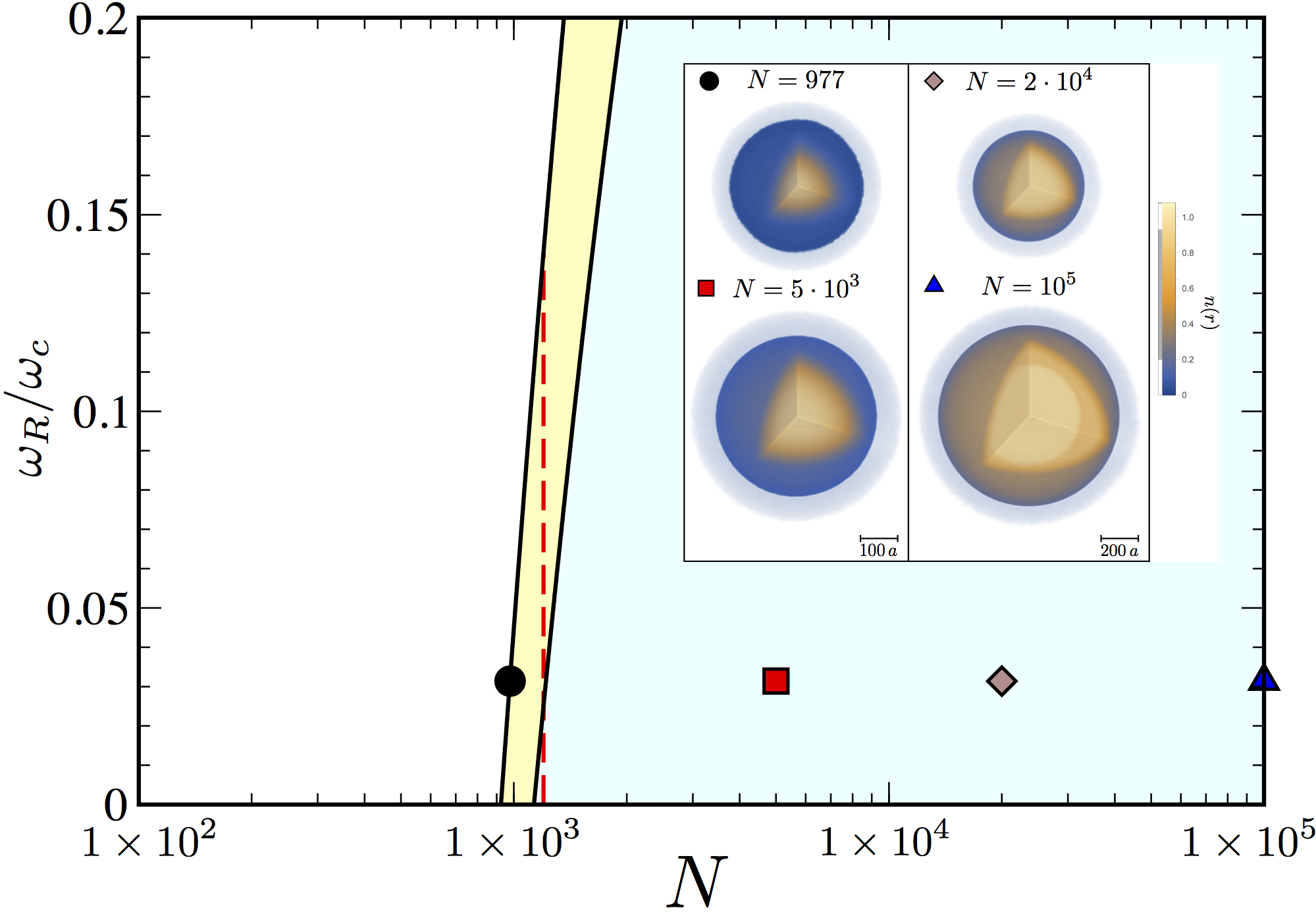}
\caption{{\bf Stability diagram of the droplet phase.} 
We identify the phases of a Rabi-coupled Bose mixture with equal number 
of particles upon the minimization of the
energy functional $\bar{U} \big(\tilde{\sigma}_1, \tilde{\sigma}_2, \tilde{\sigma}_3\big)$
of equation \eqref{potential energy per particle, dynamics}. 
We observe three phases: a stable droplet-phase region (light green) of spherical self-bound 
droplets, a metastable droplet phase (yellow) where the energy of the 
droplet is positive and larger than a uniform background with vanishing density, 
and an unstable (white region) for small particle number $N$ or high Rabi coupling $\omega$ 
where droplet evaporate. 
Here we consider $|1+\epsilon|=0.5$ which corresponds to $\omega_c \simeq 31.8\,$kHz. 
In the inset we plot the three dimensional density profile $n(r)$ of droplets from the numerical 
solution of the Gross-Pitaevskii equation for different particle 
numbers at $\omega_R/2\pi= 1$kHz,
from the metastable region $N=977$ and gaussian density limit $N=5\cdot10^3$ 
to the Thomas-Fermi regime $N=2\cdot 10^4$ and $N=10^5$ where system 
density is roughly constant up
to a critical droplet radius.
Moving along the vertical axis, increasing the Rabi coupling, droplets become metastable and finally 
unstable. Red dashed line refers to a system of $N=1200$ particles (see Methods).
}
\label{fig2}
\end{figure}

For a finite system of $N$ of particles we define a space-time dependent 
complex field $\phi(\mathbf{r},t)$ such that $n({\bf r},t) = \big|\phi(\mathbf{r},t)\big|^2$ 
is the space-time dependent local number density, and clearly $N=\int d^3{\bf r} \ n({\bf r},t)$. 
The dynamics of $\phi(\mathbf{r},t)$ is driven by the 
following real-time effective action 
\begin{equation}
S_{\text{eff}}[\phi^*,\phi] = \int dt\, d^3\mathbf{r} \,
\bigg[i\hbar\phi^* \partial_t \phi -\frac{\hbar^2\big|\nabla \phi\big|^2}{2m} 
- \mathcal{E}_{ND}\big(|\phi|^2\big) \bigg] \, , 
\label{effective action}
\end{equation}
where $\mathcal{E}_{ND}$ is obtained 
from equation (\ref{canonical energy functional}) neglecting the imaginary 
term proportional to $[(1+\epsilon)\bar{n}]^{5/2}$.
In the inset of Fig. \ref{fig2} we plot the density profile of the stationary 
solution obtained by numerically solving the Gross-Pitaevskii equation 
associated to equation \eqref{effective action} varying the number of particles for $\omega_R/2\pi = 1$ kHz. 
The solution indeed corresponds to a self-bound spherical droplet whose 
radial width increases by increasing the number of atoms. 
For a very large number of atoms, the plateau of the density profile 
approaches the thermodynamic density given by equation \eqref{trova-n}.
Instead, for a small number of atoms the self-bound droplet does not exist.
%%%%%%%%%%%%%%%%%%%%%%%%%%%%%%%%%%%%%%%%%%%%%%
%%%%%%%%%%%%%%%%%%%%%%%%%%%%%%%%%%%%%%%%%%%%%%%%%%
For small atom numbers one can model the droplet by using a Gaussian wavefunction 
\begin{equation}
\tilde{\phi}(\tilde{\mathbf{r}},\tilde{t}) = \frac{\sqrt{\tilde{N}}}{\pi^{3/4}
\sqrt{\tilde{\sigma}_1\tilde{\sigma}_2\tilde{\sigma}_3}}
\prod_{i=1}^3\exp\bigg[-\frac{\tilde{x}_i^2}{2\tilde{\sigma}_i^2} + 
i \tilde{\beta}_i \tilde{x}_i^2 \bigg]
\label{gaussian time dependent}
\end{equation}
where $\tilde{\sigma}_i(t)$ and $\tilde{\beta}_i(t)$ are time-dependent 
variational parameters rescaled in units of $a$. Here we set 
$\mathbf{r}=a\, \tilde{\mathbf{r}}$ and $\big|\phi\big| = \sqrt{n} \big|\tilde{\phi}\big|$.  
The normalization condition then becomes 
$\int d^3 \mathbf{\tilde{r}} \big|\tilde{\phi}\big|^2 = \tilde{N}$ 
where the particle number is $N = \tilde{N}\bar{n}$.
By inserting equation \eqref{gaussian time dependent} in 
the rescaled version of equation \eqref{effective action}, 
one gets six Euler-Lagrange equations for the parameters 
$\lbrace\tilde{\sigma}_i,\tilde{\beta}_i\rbrace_i$, 
i.e $\tilde{\beta}_i =\dot{\tilde{\sigma}}_i/2\tilde{\sigma}_i$ 
and $\ddot{\tilde{\sigma}}_i =  -\partial\bar{U}
\big(\tilde{\sigma}_1, \tilde{\sigma}_2, \tilde{\sigma}_3\big)/\partial 
\tilde{\sigma}_i$. 
$\bar{U} \big(\tilde{\sigma}_1, \tilde{\sigma}_2, \tilde{\sigma}_3\big)$\cite{Perez-garcia1997} is a variational
energy functional which is function of the width of the droplet only (see Methods).
The variational stability diagram of the droplet phase is illustrated in Fig. \ref{fig2}.
Upon increasing the atom number droplets stabilize. For small particle numbers
we find a metastable region where $\bar{U} \big(\tilde{\sigma}_1, \tilde{\sigma}_2, \tilde{\sigma}_3\big)$
has a local minimum with positive energy, the global minimum corresponding to zero energy for 
a dispersed gas with zero density. Interestingly, 
tuning the Rabi coupling to large values, as shown with the red dashed line for $N=1200$ particles in 
Fig. \ref{fig2}, we move into the unstable phase. 
Therefore, differently from dipolar gases \cite{Blakie2016_R} or bosonic mixtures with attractive 
inter-species interactions \cite{Petrov2015}, where transition to the instability is driven by interactions, 
here, a direct coupling between the two components serves as an additional tunable knob 
to cross from a stable into an unstable phase.

The low-energy collective excitations of the self-bound droplet are investigated 
by solving the eigenvalues problem for the Hessian matrix of effective 
potential energy in equation \eqref{potential energy per particle, dynamics}. 
From the form of the variational ansatz we naturally describe 
the monopole (breathing) mode of frequency $\omega_M$ and the quadrupole 
mode of frequency $\omega_Q$.
%%%%%%%%%%%%%%%%%%%%%%%%%%%%%%%%%%%%%%%%%%%%%
\begin{figure}
\centering
\includegraphics[width=\textwidth,clip=]{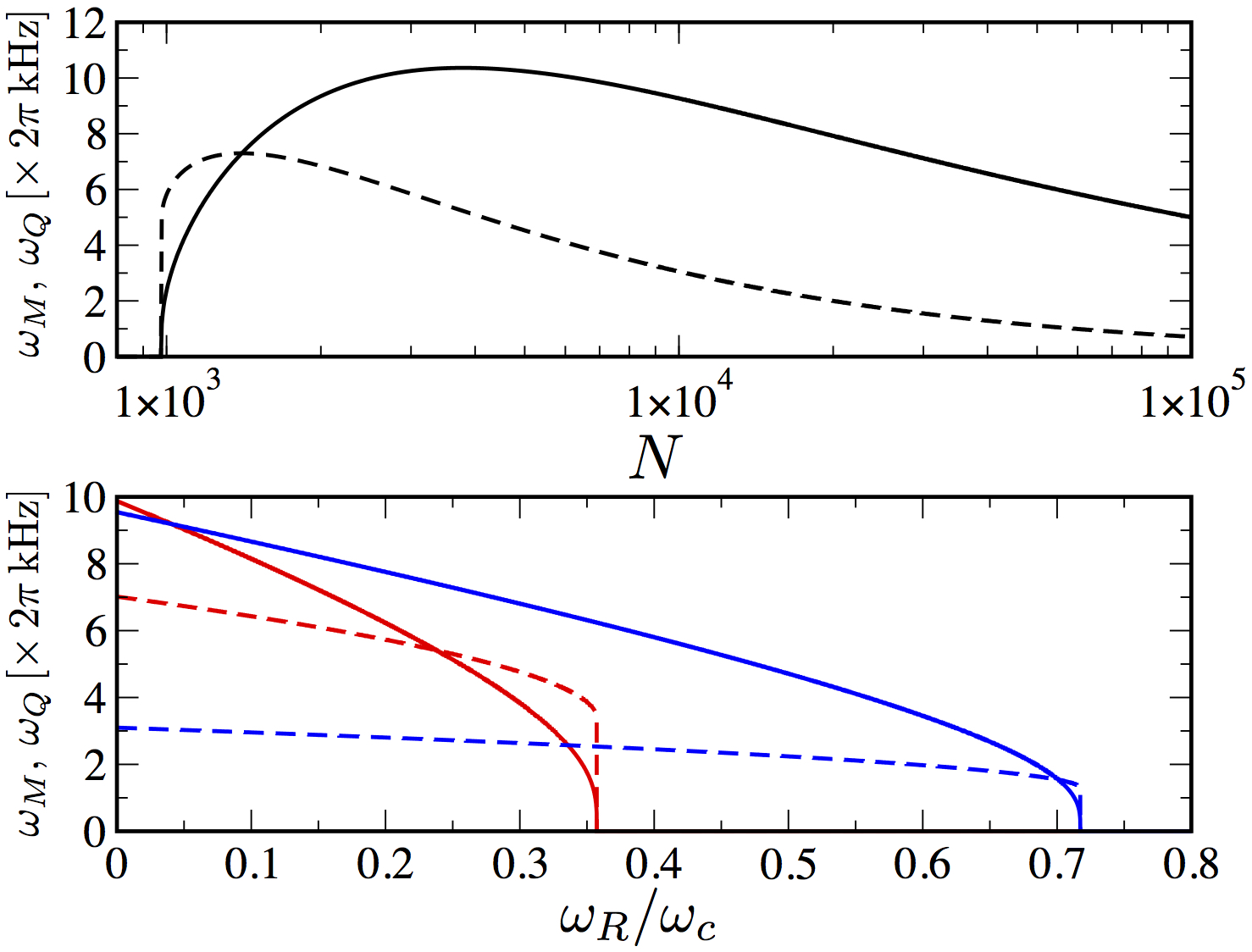}
\caption{{\bf Collective excitations of droplets.} Monopole (breathing) mode frequency $\omega_M$ (solid) 
and quadrupole mode frequency $\omega_Q$ (dashed) with $\big|1+\epsilon\big| = 0.5$. 
Upper panel: frequencies as a function of particle number and $\omega_R/2\pi= 1\,\text{kHz}$. 
Below $N\simeq 977$ the droplet becomes unstable. 
Lower panel: frequencies as a function of Rabi coupling 
for $N= 2\cdot10^3$ (red), and $N= 10^5$ (blue). 
The critical Rabi frequency occurs at $\omega_c/2\pi = 31.8$ kHz.}
\label{fig3}
\end{figure}
%%%%%%%%%%%%%%%%%%%%%%%%%%%%%%%%%%%%%%%%%%%%%
The upper panel of Fig. \ref{fig3} displays monopole and quadrupole frequencies as a 
function of the number $N$ of atoms in the droplet, fixing Rabi coupling 
and scattering lengths. The lower panel of Fig. \ref{fig3} 
reports the collective frequencies as a function of the Rabi coupling and 
two different values of $N$. Both frequencies go to zero at the Rabi 
coupling above which the droplet evaporates. 

The experimental observation of a droplet phase with Rabi coupled internal states
is within experimental reach. A promising candidate is a gas of $^{39}$K atoms loaded in 
hyperfine states $\ket{F=1, m_F = 0}$ and $\ket{F=1,m_F=-1}$. 
The narrow Feshbach resonance at 
$B \simeq 54.5$ G for collisions between atoms in $\ket{1,0}$, 
allows to tune intra-component scattering length to equal values 
to the intra-component one for the state $\ket{1,-1}$, 
then $a_1 = a_2 \simeq 40$ $a_0$, where $a_0$ is the Bohr radius 
\cite{D'Errico2007,Lysebo2010}. 
The corresponding inter-component scattering length is $a_{12} \simeq -60a_0$, 
which  gives $\epsilon \simeq -1.5$ . 
For a Rabi coupling frequencies of the order of $\omega_R/2\pi= 1$ kHz  \cite{Oberthaler2011}. 
and $N=10^5$ particles, we predict a droplet with a FWHM $\simeq 1.45$ $\mu$m.

\section*{Discussion}

We derived the beyond-mean-field grand potential of a Rabi-coupled bosonic mixture
within the formalism of functional integration, and performing regularization of 
divergent Gaussian fluctuations. 
In the small Rabi-coupling regime we also obtained an analytical expression 
for the internal energy of the system. In the case of attractive inter-particle scattering length 
we have shown how the Gaussian terms of the internal energy help 
to stabilize the system against the collapse and that, for a finite number of atoms, 
a self-bound droplet is produced. 
Rabi coupling works as an additional tool to tune the stability properties of
the droplet, inducing an energetic instability for large inter-component couplings.
The evaporation of the droplet is also signaled by both the breathing and quadruple modes which
vanish at a critical Rabi coupling.
Notably, our predictions provide a benchmark for 
experimental observations of Rabi-coupled self-bound droplets in current experiments.

\section*{Methods}
\subsection*{Quantum fluctuations and equation of state}
The inverse propagator introduced in equation (\ref{S2 fourier}) is defined by:
\begin{equation}
-\hbar \mathbb{M}(\mathbf{q},\omega_n) = 
\begin{pmatrix}
-\hbar \mathcal{G}^{-1}(\mathbf{q},\omega_n) & \hbar\Sigma_{12} \\
\hbar\Sigma_{12} & -\hbar \mathcal{G}^{-1}(\mathbf{q},\omega_n) \\
\end{pmatrix} \;
\label{propagator}
\end{equation}
(see e.g. Refs.\cite{Stoof2009Ultracold,armaitis2015}), where:
\begin{equation}
-\hbar \mathcal{G}^{-1} = 
\begin{pmatrix}
-i\hbar\omega_n + h_q & g v/\sqrt{2} \\
 g v/\sqrt{2} & i\hbar\omega_n + h_q \\
\end{pmatrix}
\;,
\label{green function 1p}
\end{equation}
is the single component inverse propagator, 
with $h_q = \varepsilon_q + g v^2 + g_{12}v^2/2 - \mu$ and 
$\varepsilon_q = \hbar^2 q^2/2m$.
The off-diagonal blocks of $\mathbb{M}(\mathbf{q},\omega_n)$ 
are given by
\begin{equation}
\hbar \Sigma_{12}= 
\begin{pmatrix}
g_{12}v^2/2 - \hbar\omega_R &  g_{12}v^2/2 \\
g_{12}v^2/2 & g_{12}v^2/2- \hbar\omega_R \\
\end{pmatrix} \;.
\label{self-energy}
\end{equation}
describes the inter-component coupling.

\subsection*{Variational and numerical analysis.}  
The equation for $\tilde{\sigma}_i$ 
is the classical equation of motion for a particle of coordinates 
$\mathbf{\tilde{\sigma}}=\big(\tilde{\sigma}_1, \tilde{\sigma}_2, 
\tilde{\sigma}_3\big)^T$ 
moving in an effective potential given by the derivative of the potential 
energy per particle:
\begin{equation}
\begin{aligned}
\bar{U}\big(\tilde{\sigma}\big) &= \frac{1}{2}\sum_{i=1}^3 \frac{1}
{2\tilde{\sigma}_i^2} 
- \frac{\big| 1+\epsilon\big|\;N}{2\sqrt{2\pi}\big(\tilde{\sigma}_1 \tilde{\sigma}_2 
\tilde{\sigma}_3\big)} 
+ \alpha \frac{(1+|\epsilon |)^{5/2}\; N^{3/2}}{\big( \tilde{\sigma}_1 \tilde{\sigma}_2 
\tilde{\sigma}_3\big)^{3/2}} \\
& \qquad \qquad \qquad \qquad \qquad + \gamma \frac{(1+|\epsilon |)^{3/2}\;
\bar{\omega}_R\;N^{1/2}}{\big(\tilde{\sigma}_1 
\tilde{\sigma}_2 \tilde{\sigma}_3\big)^{1/2}}
\end{aligned}
\label{potential energy per particle, dynamics}
\end{equation}
where $\alpha = \frac{128}{75\sqrt{5}\pi^{7/4}}$ 
and $\gamma = \frac{112}{9\sqrt{3}\pi^{5/4}}$.

The energy per particle of the ground state is simply 
$\bar{E}_{\text{gs}}/N = \bar{U}(\tilde{\sigma}_m)$ 
where $\tilde{\sigma}_m$ is the minimum of the effective potential energy. 
In absence of an external trapping, the system preserves its spherical 
symmetry, i.e. the critical point of the effective potential 
in equation \eqref{potential energy per particle, dynamics} is for
$\tilde{\sigma}_{m1} = \tilde{\sigma}_{m2} = \tilde{\sigma}_{m3}$. 
The time dependence of $\tilde{\beta}_i$ 
is completely determined by the one of $\tilde{\sigma}_i$ \cite{Perez-garcia1997}. 

\begin{figure}
\centering
\includegraphics[width=\textwidth,clip=]{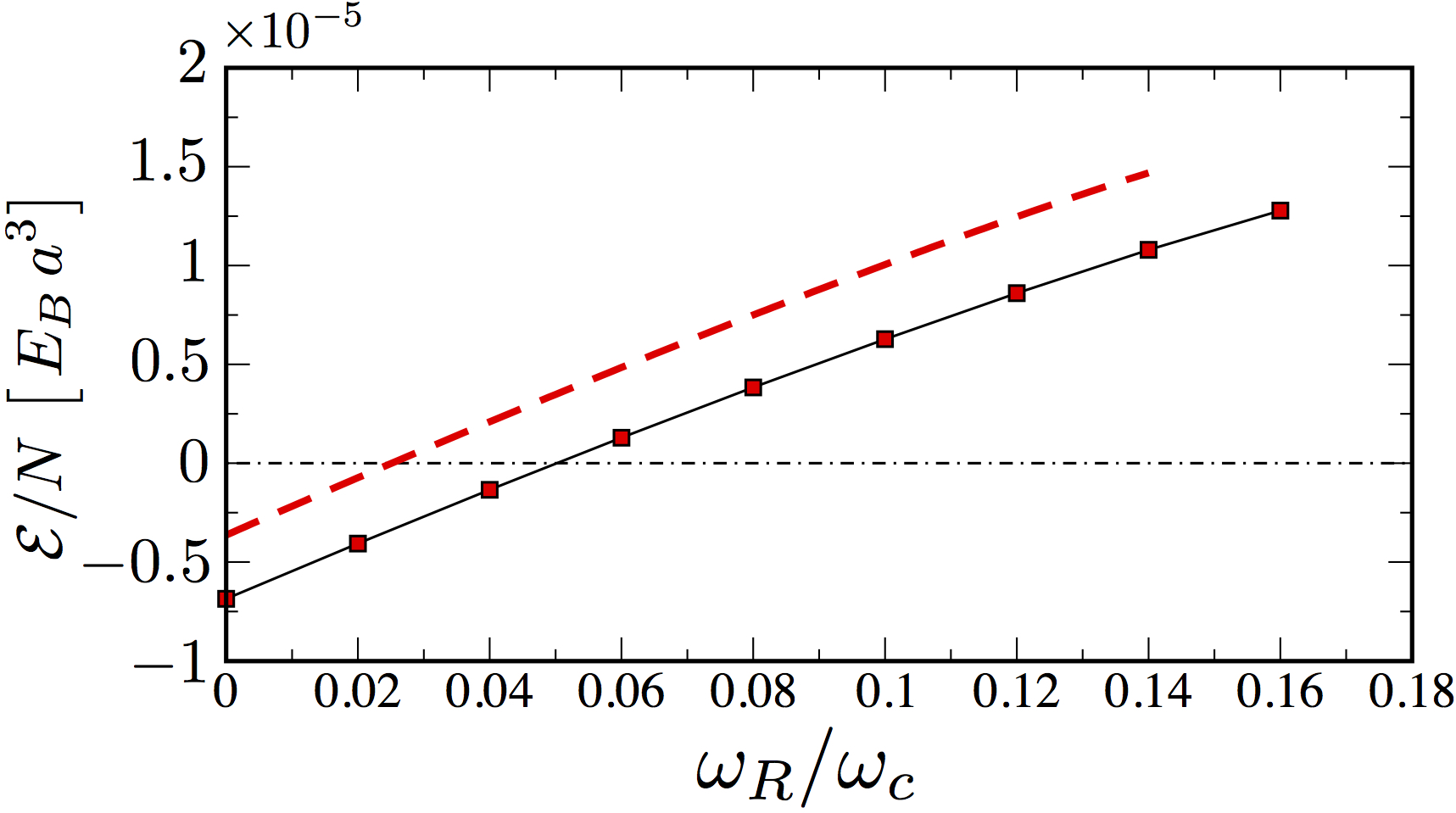}
\caption{{\bf Energetic instability of droplets.}
Energy per particle of a system of $N=1200$ particles as a function of the Rabi coupling along the vertical
line of Fig. \ref{fig2}. Red dashed line: Variational energy from equation (\ref{potential energy per particle, dynamics}).
Squared dots: Energy per particle from the numerical solution of the Gross-Pitaevskii equation. 
Increasing the Rabi coupling to values larger than $\omega\simeq 0.16\, \omega_c$ the \textit{metastable} droplet
evaporates.}
\label{fig4}
\end{figure}

Fig. \ref{fig4} shows the energy per particle of the self-bound 
droplet: the numerical approach is in reasonable agreement 
with the variational one based on equation \eqref{gaussian time dependent}. 
Remarkably, above a critical Rabi frequency the internal energy of the droplet 
becomes positive, signaling that the droplet goes in a metastable configuration. 
Moreover, at a a slightly larger critical Rabi frequency the droplet evaporates. 

\noindent \textbf{Data availability.}  Data are available upon request. Requests should be addressed to either author.

\newpage

\bibliography{References}

\begin{thebibliography}{99}

\bibitem{Leggett2001} A.J. Leggett, Rev. Mod. Phys. \textbf{73}, 307 (2001).

\bibitem{Hall1998} D.S. Hall, M.R. Matthews, J.R. Ensher, C.E. Wieman and E. A. Cornell, 
Phys. Rev. Lett. \textbf{81}, 1539 (1998).

\bibitem{Hall1998_2} D.S. Hall, M.R. Matthews, C.E. Wieman and E.A. Cornell, 
Phys. Rev. Lett. \textbf{81}, 1543 (1998). 

\bibitem{Carusotto2017} S. Butera, P. $\ddot{\text{O}}$hberg, I. Carusotto, 
arXiv1702.07533v1 (2017).

\bibitem{Larre2013} P.E. Larre and N. Pavloff, \textit{EPL} \textbf{103}, 60001 (2013).

\bibitem{Dalibard2011} J. Dalibard, F. Gerbier, G. Juzeliunas, 
A. Gostauto and P. $\ddot{\text{O}}$hberg, \textit{Rev. Mod. Phys.} \textbf{83}, 
1523 (2011).

\bibitem{Mottonen2009} V. Pietil$\ddot{\text{a}}$ and M. M$\ddot{\text{o}}$tt$\ddot{\text{o}}$nen, \textit{Phys. Rev. Lett.} 
\textbf{102}, 
080403 (2009).

\bibitem{Mottonen2014} M. W. Ray, E. Ruokokoski, S. Kandel, 
M. M$\ddot{\text{o}}$tt$\ddot{\text{o}}$nen and D. S. Hall, 
\textit{Nature} \textbf{505}, 657 (2014).

\bibitem{Merkl2008} Merkl, M., F. E. Zimmer, 
G. Juzeli$\bar{\text{u}}$nas, and P. $\ddot{\text{O}}$hberg, 
\textit{Europhys. Lett.} \textbf{83}, 54002 (2008).

\bibitem{Song2009} Jian-Jun Song and Bradley A. Foreman, \textit{Phys. Rev. A}
\textbf{80}, 045602 (2009).

\bibitem{Li2012} Y. Li, L. P. Pitaevskii and  S. Stringari, 
\textit{Phys. Rev. Lett.} \textbf{108}, 225301 (2012).

\bibitem{Martone2012} G. I. Martone, Y. Li, L. P. Pitaevskii, and 
S. Stringari, \textit{Phys. Rev. A} \textbf{86} 063621 (2012).

\bibitem{Gross11}
C. Gross, et al., 
\textit{Nature} \textbf{480}, 219 (2011).

\bibitem{Lucke11}
B. L\"ucke, et al.,
\textit{Science} \textbf{334}, 773 (2011).

\bibitem{Macri16}
T. Macr\`i, A. Smerzi and L. Pezz\`e,
\textit{Phys. Rev. A} \textbf{94}, 010102(R) (2016).

\bibitem{Fukuhara11}
T. Fukuhara, et al.
\textit{Nature} \textbf{502} 7469, 76 (13) (2013).

\bibitem{Schauss14}
P. Schauß, et al.
\textit{Science} \textbf{347} (6229), 1455-1458 (2014).

\bibitem{Zeiher16}
J. Zeiher, et al.
\textit{Nature Physics}, doi:10.1038/nphys3835 (2016).

\bibitem{Labuhn16}
H. Labuhn, et al.
\textit{Nature} \textbf{534}, 667 (2016)


\bibitem{Larsen1963} D. M. Larsen, \textit{Ann. Phys.} \textbf{24}, 89 (1963).

\bibitem{Petrov2015} D. S. Petrov, \textit{Phys. Rev. Lett.} \textbf{115}, 155302 (2015).

\bibitem{Petrov2016} D.S. Petrov, \textit{Phys. Rev. Lett.} \textbf{117}, 100401 (2016).

\bibitem{Kadau16}
H. Kadau et al., \textit{Nature} \textbf{530}, 194 (2016).

\bibitem{Wachtler2016_fil} F. W$\ddot{\text{a}}$chtler and L. Santos, 
\textit{Phys. Rev. A} \textbf{93}, 061603 (2016).

\bibitem{Schmitt16}
M. Schmitt, M. Wenzel, F. B\"ottcher, I. Ferrier-Barbut,
and T. Pfau, \textit{Nature} \textbf{539}, 259 (2016).

\bibitem{Blakie2016_R} D. Baillie, R. M. Wilson, R. N. Bisset  and 
P.B. Blakie, \textit{Phys. Rev. A} \textbf{94}, 021602(R) (2016).

\bibitem{Blakie2016} R. N. Bisset, R. M. Wilson, D. Baillie and 
P.B. Blakie, \textit{Phys. Rev. A} \textbf{94}, 033619 (2016).

\bibitem{Wachtler2016} F. W$\ddot{\text{a}}$chtler and L. Santos, 
\textit{Phys. Rev. A} \textbf{94}, 043618 (2016). 

\bibitem{armaitis2015} J. Armaitis, H. T. C. Stoof, and R. A. Duine, 
\textit{Phys. Rev. A} \textbf{91}, 043641 (2015).

\bibitem{schakel2008} A. Schakel, \textit{Boulevard of Broken Symmetries: 
Effective Field Theories of Condensed Matter} (World Scientific, Singapore, 
2008).

\bibitem{Stoof2009Ultracold} H. T. C. Stoof, D. B. M. Dickerscheid, 
and K. Gubbels, \textit{Ultracold Quantum Fields} (Springer, Dordrecht, 2009).

\bibitem{andersen2004} J. O. Andersen, \textit{Rev. Mod. Phys.} \textbf{76}, 599 (2004).

\bibitem{abad2013} M. Abad and A. Recati, \textit{Eur. Phys. J. D} \textbf{67} (7), 
148 (2013).

\bibitem{lellouch2013} S. Lellouch, T.-L. Dao, T. Koffel, and 
L. Sanchez-Palencia, \textit{Phys. Rev. A} \textbf{88}, 063646 (2013).

\bibitem{search2001} C. P. Search, A. G. Rojo, and P. R. Berman, 
\textit{Phys. Rev. A} \text{64}, 013615 (2001).

\bibitem{Tommasini2003}P. Tommasini, E. J. V. de Passos, 
A. F. R. de Toledo Piza, M. S. Hussein, and E. Timmermans, 
\textit{Phys. Rev. A} \textbf{67}, 023606 (2003).

\bibitem{Salasnich2016} L. Salasnich and F. Toigo, \textit{Phys. Rep.}
\textbf{640}, 1 (2016).

\bibitem{Fetter} A. L. Fetter and J. D. Walecka, 
\textit{Quantum Theory of Many-Particle Systems} (McGraw-Hill, Boston, 1971).

\bibitem{Diener2008} R. B. Diener, R. Sensarma and M. Randeria, 
\textit{Phys. Rev. A} \textbf{77}, 023626 (2008). 


\bibitem{Perez-garcia1997} V.M. Pérez-Garcia, H. Michinel, 
J. I. Cirac, M. Lewenstein, and P. Zoller, \textit{Phys. Rev.
Lett.} \textbf{77}, 5320 (1996).

\bibitem{D'Errico2007} C. D'Errico et al., \textit{New J. Phys.} \textbf{9}, 223 (2007).

\bibitem{Lysebo2010} M. Lysebo and L. Veseth, \textit{Phys. Rev. A} \textbf{81}, 
032702 (2010).

\bibitem{Oberthaler2011} E. Nicklas et al.,
\textit{Phys. Rev. Lett.} \textbf{107}, 193001 (2011).

\end{thebibliography}

\section*{Acknowledgements}
The authors acknowledge discussions with A. Recati and F. Toigo and F. Minardi for useful correspondence. 
T.M. acknowledges CNPq for support through Bolsa de produtividade em Pesquisa n. $311079/2015-6$ 
and the hospitality of the Physics Department
of the University of Padova.

\section*{Author contributions statement}
L.S. conceived the work.
A.C. and T.M. derived and analysed the results under the supervision of L.S. 
F.G.B. contributed in the derivation and analysis of the renormalized grand potential with Rabi coupling.
A.C., T.M. and L.S. wrote and reviewed the manuscript. 

\section*{Additional information}
\textbf{Competing financial interests:} The authors declare no competing  financial interests.

\end{document}